\documentclass[12pt]{article}
\usepackage{graphics}
\input epsf

\newcommand{\ev}[1]{\left\langle #1 \right\rangle}
\newcommand{\abs}[1]{\left| #1 \right|}
\newcommand{\binom}[2]{\left(\,
  \begin{array}{c}
    #1 \\
    #2
  \end{array}
  \,\right)}
\newcommand{\pfrac}[2]{\left(\frac{#1}{#2}\right)}
\newcommand{\be}{\begin{eqnarray}}
\newcommand{\ben}{\begin{eqnarray}\nonumber}
\newcommand{\ee}{\end{eqnarray}}

\begin{document}
\title{
\vskip 1cm
Ionic Binding in a Susy Background}
\author{L. Clavelli\footnote{lclavell@bama.ua.edu} \quad and T. Lovorn\footnote{tflovorn@bama.ua.edu}\\
Department of Physics and Astronomy\\
University of Alabama\\
Tuscaloosa AL 35487\\ }
\maketitle
\begin{abstract}
     From string theory and the observation of a positive vacuum energy in
our universe it seems inevitable that there will eventually be a phase transition
to an exactly supersymmetric (susy) universe.  In this phase there will be an 
effective weakening of the Pauli principle due to fermi-bose degeneracy.  
As a consequence molecular binding will be significantly affected.  
We make some general comments on susy molecules and perform a variational 
principle estimate of ionic binding energies.
\end{abstract}

\section{\bf Introduction}
\setcounter{equation}{0}
    Very little is known at present about the behavior of bulk
matter in a supersymmetric background.  Since supersymmetry is
a prominent feature of string vacua a prime question is whether
or not the evolution of intelligent supersymmetric life forms 
would be possible
in such a background.  If not, since we could hardly expect to be making
observations in a universe which did not support life, observer
bias would offer some type of explanation for why our universe has,
or seems to have, a broken supersymmetry (susy).  In this sense
a fuller understanding why our universe is as it is might require an
exploration of alternative universes.  

A supersymmetric
universe, at least in flat space, would be expected to have a lower
vacuum energy than the observed positive vacuum energy of our
current universe.  Therefore, from the string theory point of view, 
the current universe  is intrinsically unstable \cite{Giddings} to a 
future exactly supersymmetric universe with, possibly, zero vacuum energy.  A first 
phenomenological glance at the properties of such a universe has been already 
undertaken \cite{future}.  As discussed there the transition will begin
with the quantum nucleation of a bubble of true (susy) vacuum of radius sufficiently
large to allow bubble growth.  It is likely that this critical size is of galactic
dimensions \cite{Frampton}.  As the scale of the universe expands, the probability 
per unit time for the nucleation of a critically sized bubble will grow without limit 
until the transition occurs.  If the current acceleration continues until all matter
is trapped in white dwarfs, neutron stars, and black holes, the eventual phase transition
will lead to a rapid series of gamma ray bursts \cite{CK}
followed by collapse of all matter into susy black holes \cite{collapse}.  On the
other hand, if the transition occurs soon enough that earth-like planets are
still orbiting stars, susy planets will be formed and it is possible,
depending on susy molecular physics, that susy life forms will evolve
\cite{future}.  The modest goal of the current paper, however, is solely 
a study of the properties of ionically bound molecules assuming susy atoms
come into existence in the form of small, heavy nuclei surrounded by
an electron/selectron cloud.
     
     In our broken susy universe, the structure of matter is dominated
by the Pauli principle which forces bound fermions into high energy
levels.  After a transition to the exact susy universe, these
excited fermions will convert in pairs to their degenerate scalar
partners through gluino or photino exchange:
\be
      f f \rightarrow {\tilde f} {\tilde f} .
\label{pairconversion}
\ee

   The scalars, $\tilde f$, not being constrained by the Pauli principle, will fall into the ground state wave function via, for example, photon emission.  Thus, in exact susy, there will be no more than two fermions of each type in each bound state.  The remaining baryon and lepton number will be filled out by sbaryons and sleptons with all stable nuclei and atoms containing particles in s-waves only.  Where possible without confusion, we
will refer generically to fermionic and scalar electrons as electrons and to fermionic and scalar nucleons as nucleons.  In exact susy all particles and
all bound states are members of degenerate susy multiplets.

   Our variational approach allows an approximate but analytic solution for the ion energies, mean
radii, and ionic binding energies.
For the sake of numerical comparison with our world, we evaluate our formulae in the
case that the masses of the degenerate susy particles will be the masses of the standard model 
particles in the broken susy 
world.  This is approximately what would be expected if the ground state of the universe 
corresponded to one of the higgs extensions of the minimal supersymmetric standard model (MSSM) with all susy breaking parameters 
set to zero\footnote{S. Nasri, private communication}. 
In these models  
the coupling constants would also be expected to be approximately as they are in the broken susy world
with small deviations given by the running of the couplings over a slightly larger domain of
exact susy. 
Other models for susy breaking in our world, such as gauge mediated models, 
could lead to similar results as the susy breaking is turned off.  In the model of radiative
breaking of electroweak symmetry, the symmetry breakdown of the standard model is due to a
squared Higgs mass running to negative values.  This could continue to be true if exact
supersymmetry extends into the low energy domain 
  
One could also entertain alternative possibilities for the ground state of the universe such as the supersymmetric AdS/CFT model or one of the five flat space superstring theories.  These
might qualitatively share some of the features we investigate here.  We give analytic results for the
energies and mean radii, so the consequences of other assumptions about the degenerate susy multiplet masses and charges can be readily evaluated. 

   In section II we approximate the multi-electron wave function as products of
hydrogenic wave functions.  This wave function is improved by the
variational method to take into account screening in section III.  In section IV
we calculate the ionic bonding of some diatomic molecules that are prominent
in our broken susy world and in section V we treat ionic binding of the susy
water molecule.  We conclude with a summary and some discussion in section VI.

\section{\bf Susy ions}
\setcounter{equation}{0}
Working in natural units, $\hbar = c = 1$,
the Hamiltonian for the system of nuclear charge $Z$ and $N$ electrons is given by:
\begin{equation}\label{H1}
H=-\frac{1}{2m}\sum_{i=1}^N \nabla_i^2-Ze^2\sum_{i=1}^N \frac{1}{r_i}+e^2 \sum_{i
< j}^N \frac{1}{\left|\mathbf{r_i}-\mathbf{r_j}\right|}
\end{equation}
with $e^2$ being the fine structure constant, $\alpha \approx 1/137$, and $m$ being
the electron mass.

From the variational principle, an upper bound on the ground state
energy is given by the expectation value of the Hamiltonian for any
trial wavefunction.
\begin{equation}
E_{gs}=\left\langle\psi|H|\psi\right\rangle \quad .
\end{equation}

The ground state wavefunction for the hydrogen atom  
is given by:
\begin{displaymath}
\psi_{100}=\sqrt{\frac{(m e^2 Z)^3}{\pi}}e^{- m e^2 Zr} \quad .
\end{displaymath}
We might take the trial wavefunction for the general ion to be the product
of this ground state wavefunction for each electron.
\begin{equation}
\psi=\left(\frac{(m e^2 Z)^3}{\pi}\right)^{N/2}e^{-m e^2 Z\sum r_i} \quad .
\end{equation}

This is an eigenfunction of the first two terms in eq.\,\ref{H1}
with eigenvalue
\begin{equation}
\ev{H}_{Hydrogenic}=- R_\infty Z^2N
\label{hydrogenic}
\end{equation}

where $R_\infty$ is the Rydberg constant
\be
     R_\infty = m e^4/2 = 13.6 \displaystyle{eV}.
\ee
This wave function is exact for the hydrogenic atoms ($N=1, Z$ arbitrary)
ignoring nuclear effects and relativistic corrections.
For $N > 2$ fermionic electrons, this would not be expected to be a good 
approximation since only two electrons can be in this $1s$ wave function.
However, in the exact susy world, all of the charged leptons can be in the
ground state wave function with as many of them as necessary becoming bosons
via the pair conversion process of eq.\,\ref{pairconversion}.  Note that
the conversion process occurs generically in susy and does not require
$R$ parity violation.

Now, we need to calculate the expectation value of the last term,
corresponding to the repulsion between the electrons
\begin{equation}
V_{ee} \equiv e^2 \sum_{i<j}^N
\frac{1}{\left|\mathbf{r_i}-\mathbf{r_j}\right|} \quad .
\end{equation}

We calculate $\ev{V_{ee}}$ directly by integration.
\begin{equation}
\ev{V_{ee}} = e^2 \int{\psi^2\left(\sum_{i>j}^N
\frac{1}{\left|\mathbf{r_i}-\mathbf{r_j}\right|}\right)
d\mathbf{r_1} d\mathbf{r_2} d\mathbf{r_3} \dots d\mathbf{r_N}} \quad .
\end{equation}

The number of terms in the sum is equal to $\binom{N}{2}$. By
symmetry, this integral can be rewritten and simplified.

\begin{equation}
\ev{V_{ee}} = e^2 \binom{N}{2}\left(\frac{(me^2 Z)^3}{\pi}\right)^2
\int{e^{-2(me^2 Z)(r_1+r_2)}
\left(\frac{1}{\left|\mathbf{r_1}-\mathbf{r_2}\right|}\right)
d\mathbf{r_1} d\mathbf{r_2}} \quad .
\end{equation}

To evaluate this integral, we set the coordinates relative to
$\mathbf{r_1}$, so that the angle between $\mathbf{r_1}$ and
$\mathbf{r_2}$ is simply $\theta_2$.  By the law of cosines,
\begin{displaymath}
\left|\mathbf{r_1}-\mathbf{r_2}\right| =
\sqrt{r_1^2+r_2^2-2r_1r_2\cos{\theta_2}} \quad .
\end{displaymath}
We start by evaluating with respect to $\mathbf{r_2}$, defining a
new integral:
\begin{equation}
I_2 \equiv \int{e^{-2(me^2 Z)r_2}
\frac{1}{\sqrt{r_1^2+r_2^2-2r_1r_2\cos{\theta_2}}}
r_2^2\sin{\theta_2}dr_2d\theta_2d\phi_2} \quad .
\end{equation}

The $\phi_2$ integral is just $2\pi$.  The $\theta_2$ integral is a
bit more involved:
\begin{eqnarray}
\int_0^{\pi}{\frac{\sin{\theta_2}}
{\sqrt{r_1^2+r_2^2-2r_1r_2\cos{\theta_2}}}d\theta_2} & = &
\left[\frac{-1}{r_1r_2}\sqrt{r_1^2+r_2^2-2r_1r_2\cos{\theta_2}}\right]_0^{\pi} \nonumber \\
& = & \frac{1}{r_1r_2}\left(r_1+r_2-\abs{r_1-r_2}\right) \quad .
\end{eqnarray}
This is $2/r_1$ for $r_2<r_1$ and $2/r_2$ for $r_2 > r_1$.

Inserting these back into $I_2$, we get the following:
\begin{eqnarray}
I_2 & = & 4\pi\left(\frac{1}{r_1}\int_0^{r_1}{e^{-2(me^2Z)r_2}r_2^2dr_2} +
\int_{r_1}^{\infty}{e^{-2(me^2Z)r_2}r_2dr_2}\right) \\
& = & \frac{\pi}{(me^2 Z)^3r_1}\left(1-\left(1+me^2 Zr_1\right)e^{2me^2 Z r_1}\right) \quad .
\end{eqnarray}

Returning to $\ev{V_{ee}}$, we now have:
\begin{eqnarray}
\ev{V_{ee}} 
 & = & e^2 \binom{N}{2}4(me^2 Z)^3\int_0^{\infty}{e^{-2 me^2 Z r_1}r_1
\left(1-\left(1+ me^2 Z r_1\right)e^{-2 me^2 Z r_1}\right)dr_1}\\\nonumber
 & = & \binom{N}{2}\frac{5 R_\infty Z}{4} \quad .
\end{eqnarray}

Including the hydrogen-like term from eq.\,\ref{hydrogenic}, we have
\be
    \ev{H} = - R_\infty Z N \left( Z - \frac{5}{8}(N-1)\right) \quad .
\label{E0}
\ee

By the variational principle this gives an upper limit to the ground state
energy or a lower limit to the ground state binding energy. 

If $N=Z$,
\be
    \ev{H} = - \frac{R_\infty}{8}(3 Z^3 + 5 Z^2) .
\ee

\section{A Better Approximation}
We can improve this upper bound anticipating that electron shielding
of the nuclear charge
will reduce the coefficient in the wave function exponential.
With $Z_s$ a dimensionfull
parameter to be determined by energy minimization we take the 
normalized trial wavefunction
for the $N$ electrons to be:
\begin{equation}
\psi = \prod_{i=1}^{N}\left( \sqrt{\frac{Z_s^3}{\pi}} e^{-Z_s r_i}\right)\quad .
\label{wavefn}
\end{equation}

Proceeding as in section 2, we can write the expectation
values
 
\begin{eqnarray}
\ev{\frac{1}{r_i}}  = 4\pi\int_0^{\infty}{\pfrac{Z_s^3}{\pi}e^{-2Z_sr}r dr} = Z_s \quad,
\end{eqnarray}

\be
\ev{\nabla_i^2} = \ev{\frac{d^2}{dr_i^2}+\frac{2}{r_i}\frac{d}{dr_i}} = -Z_s^2 \quad,
\ee
and
\be
\ev{\frac{1}{\abs{\vec{r}_i-\vec{r}_j}}} = \frac{5}{8} Z_s\quad .
\ee

Thus
\be
  \ev{H} = N \frac{Z_s^2}{2m} - Z e^2 Z_s N + \frac{N(N-1)}{2} e^2 \frac{5}{8} Z_s \quad .
\ee

This is a minimum at

\be
        Z_s = m e^2 \left( Z - \frac{5}{16}(N-1) \right) \quad .
\ee

Since $Z_s$ must be positive in order to have a normalizeable wave function, this equation is only meaningful for 
\be
N < \frac{16}{5}Z + 1 \quad .  
\label{Nlimit}
\ee
Putting this back into the expression for $\ev{H}$ gives
\be
    \ev{H} = E(N,Z) = - R_\infty N \left(Z- \frac{5}{16}(N-1)\right)^2 \quad .
\label{ENZ}
\ee

Comparing with eq.\,\ref{E0}, we see that the ground state energy estimate is reduced by the improved wave function providing a decreased upper limit to the actual energy
and an increased lower limit, $-E(N,Z)$, to the actual binding energy.  According to the variational principle, further improvements in the
trial wave function will yield a binding energy approximation approaching the exact solution from below.  This variational result was known shortly after 
Schroedinger's original result \cite{Eckart} 
and improves the accuracy of the
theoretical energy of the helium-like atoms ($N=2, Z$ arbitrary) by 
a factor of three above the result based on treating
the electron repulsion perturbatively \cite{Pauling}.  

Beyond $N=2$ the
result is not a useful approximation in the broken susy world due to the
Pauli principle which forbids more than two electrons from occupying the
$1s$ level.  With the circumvention of the Pauli
principle in the exact susy case, we can expect comparable accuracy for
susy atoms of arbitrary $N$ as for standard helium. 

The average distance between an electron and the nucleus is
\be
  r(N,Z) =  4\pi \frac{Z_s^3}{\pi}\int_0^\infty r^3 dr e^{-2Z_s r} = \frac{3}{2 Z_s} 
= \frac{3 \alpha}{4 R_\infty} (Z - \frac{5}{16}(N-1))^{-1} \quad .
\ee
except that for $N=0$ we must put $r(N,Z)=0$. 

We see that, at fixed $Z$, this mean distance is increased over the hydrogenic 
($N=1$) case due to the electron repulsion.
The 
size of the neutral atoms ($N=Z$) decreases rapidly for increa‌sing $Z$ unlike the
case in the broken susy world as discussed below.  However, at least for 
moderate $Z$, the mean radius of the electron cloud remains large compared to
the nuclear size.

\section{Ionic Binding}

Ionic binding occurs when it is energetically advantageous to transfer
one or more electrons
from one atom to another after which the positive and negative ions strongly attract
each other.  Ionic binding, when appreciable, tends to dominate over covalent binding
which corresponds to two atoms sharing electrons equally.  In the broken susy world,
ionic binding is most prominent in the alkali-halide compounds where one electron can
be taken from an outer shell of one atom and used to form a closed shell on the other atom.  

The change in energy after 
transferring $l$ electrons/selectrons from one ion of atomic number $Z_1$ to another 
of atomic number $Z_2$ and taking
into account the subsequent Coulomb attraction is
\ben
    \Delta E = -E(N_1,Z_1)-E(N_2,Z_2)+ E(N_2+l,Z_2)+E(N_1-l,Z_1)\\ 
 -\, \frac{e^2 l^2}{r(N_1-l,Z_1)+r(N_2+l,Z_2)} \quad .
\ee
The negative of $\Delta E$ is the molecular binding energy.  This expression
assumes that the perturbation of the atomic wave functions away from
spherical symmetry can be neglected and that the interatomic distance is
rougly the sum of the radii of the two ions.
These approximations are reasonable as
a first attempt to estimate molecular properties in an
exact susy background.

Assuming only single electron transfer,
The ionic molecular binding energy for two neutral atoms of atomic numbers
$Z_1$ and $Z_2$ is
\ben
    B =  +E(Z_1,Z_1)+E(Z_2,Z_2)- E(Z_2+1,Z_2)-E(Z_1-1,Z_1)\\
 +\, \frac{e^2}{r(Z_2+1,Z_2)+r(Z_1-1,Z_1)} \quad .
\label{bindingenergy}
\ee

This can be written
\be
     B = F(Z_2) - I(Z_1) + e^2/(r_1+r_2)
\label{MBE1}
\ee
where the electron affinity of the atom of atomic number $Z_2$, 
namely the energy released when
an electron is added to the atom , is
\be
    F(Z_2) = E(Z_2,Z_2)-E(Z_2+1,Z_2)
\ee
and the ionization energy $I$ of atom $1$, the energy required to remove an
electron, is
\be
    I(Z_1) = - E(Z_1,Z_1) + E(Z_1-1,Z_1) \quad .
\ee
Which atom donates an electron and which receives is determined by maximizing 
the binding energy.  In general, the atom of lower $Z$ is the donor atom.  
In eq.\,\ref{MBE1} we have estimated the equilibrium separation of the
two ions as the sum of the mean radii since the overlapping electron clouds 
will inhibit closer approach.  Using the sum of the rms radii does not
lead to significantly different results.

     In fact, however, we find that, when the binding is positive, the
maximum binding comes when the atom of lesser $Z=Z_1$ donates all its
electrons to the other atom of greater atomic number $Z_2$.  This does
not result in a violation of eq.\,\ref{Nlimit}.

The binding energy is then
\be
    B = E(Z_1,Z_1) + E(Z_2,Z_2) - E(0,Z_1)-E(Z_2+Z_1,Z_2) +
          \frac{\alpha Z_1^2}{r(Z_1+Z_2,Z_2)}
\label{MBE}
\ee

The mean radius of the totally stripped ion, $r(0,Z)$, is taken to be zero. 
The molecular binding energy for hydrogen with an element of
atomic number $Z$ can, therefore, be expressed as
\be
   B = E(Z,Z) - E(Z+1,Z) - 13.6 \displaystyle{eV} + \alpha/r(Z+1,Z)\quad .
\ee
This is the electron affinity of the element of atomic number $Z$
minus the hydrogen ionization energy plus the negative of the Coulomb
energy of attraction of the two ions.

\begin{table}[htbp]
\begin{center}
\begin{tabular}{||l|l|ccc||}\hline
   &  &        &         &                                  \\
 Z  &  & $I_{susy}$   &   $I_{exp}$  &   $r_{susy}$  \\
\hline
\hline
  &        &       &       &          \\
1 &    H   & 13.6  &  13.6 &   0.792  \\
3 &    Li  & 23.8  &  5.4  &   0.023  \\
11&    Na  & 64.6  &  5.1  &   0.0070  \\
19&    K   & 105  &  4.3  &   0.0041  \\
37&    Rb  & 197  &  4.2  &   0.0021  \\
55&    Cs  & 289  &  3.9  &   0.0014  \\
  &        &       &       &          \\
\hline
\end{tabular}
\end{center}
\caption{ single electron ionization energies of alkali atoms in eV in the exact susy case
using the optimum wave function from eq.\,\ref{wavefn}. In the fourth
column we give the experimental ionization energies in our broken
susy world.  The fifth column gives the predicted mean radius of the
electron/selectron cloud in Angstroms. }
\label{ionizationenergies}
\end{table}

\begin{table}[htbp]
\begin{center}
\begin{tabular}{||l|l|cc||}\hline
   &    &        &                                  \\
 Z &    & $F_{susy}$   &   $r_{susy}$  \\
\hline
\hline
   &    &       &          \\
9  &  F   & 35.38 &   0.122  \\
17 &  Cl  & 146.3 &   0.066  \\
35 &  Br  & 669.4 &   0.033  \\
53 &  I   & 1571  &   0.022  \\
   &      &       &          \\
\hline
\end{tabular}
\end{center}
\caption{ electron affinities in eV and mean radii in Angstroms
of the halide atoms in the exact susy case
using the trial wave function from eq.\,\ref{wavefn}. }
\label{affinities}
\end{table}
 
\begin{table}[htbp]
\begin{center}
\begin{tabular}{||l|cccccc||}\hline
     &        &         &          &             &           &           \\
     &  H     &   Li    &   Na     &    K        &    Rb     &     Cs    \\
\hline
\hline
     &        &         &          &             &           &          \\
F    & 134    &  557    &  -.089   &  6610       &   24000   &   44900   \\
     &(5.87)  &  (5.91) &    (5.3) &  (5.07)     &   (5.0)   &   (5.1)   \\
Cl   & 93.14  &  1570   &   4140   &  -4.5       &   47000   &  106000   \\
     & (4.43) &  (4.8)  &    (4.2) &  (4.34)     &   (4.3)   &  (4.58)   \\
Br   & 113.1  &  4680   &    28100 &  41400      &   -12.7   &  139000   \\
     & (3.83) &  (4.3)  &    (3.74)&  (3.91)     &   (3.9)   &  (0.34)   \\
I    & 134.2  &  8920   &  56300   &  112000     &   94800   &   -21.0 \\
     &        &  (3.5)  &   (3.0)  & (3.31)      &   (3.3)   &  (3.56)   \\
\hline
\end{tabular}
\end{center}
\caption{Binding energy estimates in eV from eq.\,\ref{MBE} 
for the alkali halides in the exact susy case
using the trial wave function from eq.\,\ref{wavefn}.  Negative entries
correspond to combinations with no ionic binding in the susy world
although they are bound in the broken susy world.
Experimental binding energies in the broken susy case,
given in parentheses, are taken from \cite{Huber}.  
}
\label{bindingenergies}
\end{table}

    In table \ref{ionizationenergies} we tabulate the ionization energies
of the alkali atoms and
the mean atomic radii in the susy case to compare with the experimental values
in the broken susy case.  It is seen that, for large $Z$ in the susy case,
the ionization energy is significantly greater 
than the experimental values in our broken susy world as would be
expected since all the electrons and selectrons are in the $1s$ ground
state.  In the current universe, all atoms have a radius of order $1$ A$^\circ$
since, for greater Z, the increased attraction to the nucleus is balanced
by electron repulsion and by the need to put electrons into higher orbitals 
which have greater spatial
extent.  In exact susy on the contrary, as can be seen from table 
\ref{ionizationenergies}, heavier 
atoms have significantly smaller radii.  In table \ref{affinities}, we give 
the electron affinities for the halides, namely the energy released if an 
extra electron is dropped onto the neutral atom making a negatively charged
ion.  Table \ref{bindingenergies} gives the estimated binding energies for
the alkali halides in the susy case following eq.\,\ref{MBE}.  The negative
entries correspond to molecules that are not bound in the susy case although
they do form diatomic molecules in the broken susy case. 
Covalent binding
of alkali-halides is negligible in the broken susy world and 
presumably also negligible in the exact susy world at least for those
molecules that have significant positive ionic binding energy.

 Where the susy binding
energies are positive they are much greater than in broken susy and
the interatomic radii are much smaller.
    The smaller radii of the susy atoms suggests that chemical reaction rates will be much slower in the susy world, while the greater molecular binding energies suggests that the diatomic molecules which do bind will be much sturdier than in the broken susy case since the dissociation energies reach well into the
gamma ray region.  
However, these results are moot if, as suggested by
the semi-empirical mass formula \cite{future}, nuclei above oxygen 
are rare or nonexistent in a susy background.  If there were non-negligible
surviving heavy elements, due to the very large binding energies in table
\,\ref{bindingenergies}, they would rapidly steal atoms from lighter molecules.
 
It is also interesting to look at the low-lying diatomic molecules whose
susy binding energies in the current approximation are given in table 
\,\ref{low-lying}.
Unlike the case in the broken susy world,  in exact susy there is significant 
binding of helium to other low-lying elements.  Molecules with small or
negative ionic binding energy could be bound through covalent binding
which is not treated in this article.  We find no 
ionic binding in
diatomic molecules whose atoms differ by zero or one unit of atomic number
except for a few of the lighter molecules such as helium hydride which is
only marginally bound in broken susy.

\begin{table}[htbp]
\begin{center}
\begin{tabular}{||l|ccccccccc||}\hline
   &       &       &        &      &       &       &        &       &      \\
   & H     & He    & Li     &Be    & B     & C     & N      & O     &F     \\
\hline
\hline
   &       &       &        &      &       &       &        &       &      \\
H  & -1.9  &11.0   &25.1   &40.4   &56.8   &74.4   &93.2   &113.1  &134.2  \\
   & (4.48)&(.003) &(2.43) &(2.03) &(3.42) &       &($<$3.47)&(4.39) &(5.87) \\
He &       &-16.5  &27.6   &74.1   &123    &174    &227    &283    &341   \\ 
   &       &(0.0009) &     &       &       &       &       &       &      \\
Li &       &       &-57.1  &36.4   &133    &234    &338    &446    &557   \\
   &       &       &(1.4)  &       &       &       &       &(3.49) &(5.91) \\
Be &       &       &       &-137   &23.9   &190    &360    &536    &716   \\
   &       &       &       &       &       &       &       &(4.6)  &(5.8)  \\
B  &       &       &       &       &-270   &-23.2  &230    &489    &754   \\
   &       &       &        &      &(3.0)  &(4.6)  &       &(8.28) &(7.8)  \\
C  &       &       &       &       &       &-470   &-118   &240    &606   \\
   &       &       &        &      &       &(6.21) &(7.7)  &(11.0) &       \\
N  &       &       &       &       &       &       &-749   &-275   &207   \\
   &       &       &        &      &       &       &(9.76) &(6.50) &(3.5)  \\
O  &       &       &       &       &       &       &       &-1121  &-507  \\
   &       &       &        &      &       &       &       &(5.12) &       \\
F  &       &       &       &       &       &       &       &       &-1600  \\
   &       &       &       &       &       &       &       &       &(1.60) \\
\hline
\end{tabular}
\end{center}
\caption{Binding energy estimates in eV from eq.\,\ref{MBE} for the low lying 
diatomic molecules
in the exact susy case
using the trial wave function from eq.\,\ref{wavefn}.  Negative entries
correspond to unbound combinations. Experimental binding energies \cite{Huber}
in the broken susy 
world are given in parentheses for the molecule indicated by the line above.}
\label{low-lying}
\end{table}
   
In the case of homonuclear diatomic molecules we can write, following 
eq.\,\ref{bindingenergy}, the ionic binding energy due to the transfer of 
$l$ electrons from one atom of atomic number $Z$ to the other as
\ben
    B = 2 E(Z,Z) - E(Z-l,Z)-E(Z+l,Z) + \frac{l^2 \alpha}{r(Z-l,Z)+r(Z+l,Z)} \\
    = - \frac{l^2 R_\infty}{384}\left( 79 Z + 70 + \frac{400 l^2}{11 Z + 5}
        \right) \quad .
\ee
Since this is negative for arbitrary $l$, the homonuclear diatomic molecules 
have no ionic bonding for the approximate wave functions we are using.  
Although no ionic bonding is
found for these potential molecules, the negative binding energies are small
compared to those of the alkali-halides.  This suggests that with a better
approximation of the wave functions, ionic binding may occur and,
additionally, one might expect covalent binding to be effective for these molecules.

\setcounter{equation}{0}
\section{\bf Supersymmetric water}
\setcounter{equation}{0}
As a final example we consider the ionic binding of the susy water molecule.
In the current (broken susy) universe, water binds due to a transfer of an electron
from each of the two hydrogen atoms to $2p$ orbitals of the oxygen atom resulting
in a closed shell of oxygen.  Since the $p$ orbitals are in one to one correspondence
with the Cartesian coordinates, the hydrogen bonds, in zeroth approximation, 
are at right angles to each other.  Coulomb repulsion of the protons
then leads to the slightly greater observed bond angle of $105^\circ$.

In the susy case, all the electrons can be put into the spherically symmetric $1s$ wave function of oxygen, converting, if necessary, into scalar electrons via 
eq.\,\ref{pairconversion}.  Proton repulsion will then naturally lead to a linear triatomic molecule with the oxygen ion in the center.  Since this arrangement
has no molecular dipole moment, the properties of water as a solvent might be
significantly changed in susy matter.

In this linear configuration, the binding energy of a neutral atom of atomic
number $Z$ with two hydrogen atoms would be
\ben
    B = E(Z,Z)-E(Z+2,Z)-2 R_\infty + \frac{3 \alpha}{2 r(Z+2,Z)} \\
      = \frac{R_\infty}{128}\left( 11 Z^2 + 66 Z - 311 \right)\quad .
\ee
For $H_2 O$ this corresponds to a total dissociation energy of $98$ eV.  Atoms
lighter than beryllium will not have a positive ionic binding energy
to two hydrogen atoms.

\setcounter{equation}{0}
\section{\bf Conclusions}
\setcounter{equation}{0}

     The variational approach we have taken in this paper allows for an
approximate but analytic expression for the atomic energies, mean radii, and
ionic molecular binding energies as a function of the electron charge and 
mass.  We have tabulated numerical results for the case where the 
degenerate electron/selectron mass is equal to the electron mass in the
broken susy world but, obviously, other masses can be used if that becomes
theoretically motivated.  Our core assumptions are that, in the susy background
as in our world, the electron mass is non-zero but much less than the nucleon
mass, 
the nucleus is small compared to atomic scales, and atomic structure is
governed by a $U(1)$ electromagnetism.  The results of the paper are also
independent of any assumptions about atomic abundances as long as these
are non-zero.  Of course, the degenerate susy masses and the
abundances would be important parameters in discussing the transition
to exact susy and the subsequent properties of the susy world.

     Our approximate atomic wave function corresponds to putting all electrons
into a screened $1s$ wave function.  This approximation is
of limited usefulness in the broken susy world for atoms above helium 
due to the Pauli principle but
could be a good approximation in the exact susy world. 
We have neglected mixing between
atoms with the same $Z$ but different numbers of scalar electrons.  
We have computed ionic binding energies of various atoms assuming
that atoms remaining spherically symmetric is a reasonable zeroth
approximation as in the broken susy world.  We have also assumed that the
bond lengths are approximately the sum of the mean atomic radii.
This assumption is found to be reasonable in broken susy. We have treated
the atomic ions quantum mechanically but the molecular bound states
are treated classically.  A similar semi-classical calculation gives
reasonable results for the alkali-halides in the broken susy world.

     As expected, we find that the atomic sizes are quite small in general
compared to the corresponding atomic sizes in the broken susy world.  This
tends to lead to large ionic binding energies since the transfer of an
electron from one atom to another leads to a strong Coulomb attraction. 
Such molecules could be expected to be much sturdier in a susy world.
On the other hand some of the heavier alkali halides that are bound in 
the broken susy world become unbound in exact susy.  This is due to the
higher ionization energy of the heavy alkali atoms in exact susy where all
the electrons are in the $1s$ wave function.

     It is interesting to inquire whether the diatomic molecules
that are unbound in the susy world play any essential roles in the broken
susy world.  This might lead to an anthropic explanation of why our
world is not exactly supersymmetric.
In fact however, from an environmental point of view
one readily comes on reasons to welcome the absence of
a few of the molecules that are unbound in the susy universe
but have severe deleterious effects in our broken susy world.  
For example, among the low-lying diatomic molecules, sodium fluoride, NaF,
potassium chloride, KCl, and potassium fluoride, KF, are known to have
life-threatening effects in the broken susy world.
Of these, NaF and KCl, are seen in table \ref{bindingenergies} to be
unbound in the exact susy world.

     The susy water molecule is predicted to be linear rather than 
forming the familiar $105^\circ$ bond angle seen in the broken susy world. 
The resulting absence of a dipole moment could be 
significant with respect to viability
since, in the broken susy world, the anomalously large dipole moment of the
water molecule is important in making water a universal solvent
facilitating chemical reactions.

     Clearly, we have barely begun to explore the properties of bulk susy
matter and we are far from being able to make any statements about the
possibility of life in the susy world.  
In the more proximate future it should be possible to explore the effects of mixing 
between states of differing selectron numbers and to estimate covalent binding
energies as well as to seek an improved variational wave function.

{\bf Acknowledgements}

    This work was supported in part by the US Department of Energy under
grant DE-FG02-96ER-40967.  We acknowledge helpful discussions with R. Tipping.

\end{document}